\documentclass[twocolumn,aps,prl,superscriptaddress,nofootinbib]{revtex4-1}
\usepackage{graphicx}
\usepackage{subcaption}
\usepackage{color}
\graphicspath{{figures/}{fig/}}
\usepackage{amsmath}
\usepackage{amssymb}
\usepackage{bm}
\usepackage{slashed}
\usepackage{epsfig}
\usepackage{amsfonts}
\usepackage{epstopdf}
\usepackage{hyperref}
\usepackage{bbm}
\usepackage{textcomp}
\usepackage{color}

\newcommand{\sect}[1]{{\it \textbf{#1.} --- }}

\begin{document}
\title{Reconstruction of rational functions made simple}

\author{Xiao Liu}
\email{xiao.liu@physics.ox.ac.uk}
\affiliation{Rudolf Peierls Centre for Theoretical Physics, Clarendon Laboratory, Parks Road, Oxford OX1 3PU, UK}

\date{\today}

\begin{abstract}
We present a new method for the reconstruction of rational functions through finite-fields sampling that can significantly reduce the number of samples required. The method works by exploiting all the independent linear relations among target functions. Subsequently, the explicit solutions of the functions can be efficiently obtained by solving the linear system. As a first application, we utilize the method to address various examples within the context of Feynman integrals reduction. These examples demonstrate that our method can substantially improve the computational efficiency, making it useful for future computations in particle physics.
\end{abstract}

\maketitle
\allowdisplaybreaks

\sect{Introduction}
With the discovery of the Higgs boson~\cite{ATLAS:2012yve, CMS:2012qbp}, the particle physics Standard Model (SM) is complete. To further deepen our understanding of the SM and probe signals of new physics through potential deviations between theory and experiment, it is essential to compare high-precision results from both sides. Currently, experimental measurements and theoretical simulations for many important scattering processes are of comparable accuracy~\cite{Heinrich:2020ybq}. However, in many cases, it is expected that with the accumulation of experimental data at the Large Hadron Collider (LHC) and improvements in data analysis methods, experimental precision will surpass that of theory~\cite{Cepeda:2019klc}. Therefore, it is crucial for theorists to compute higher and higher order corrections in perturbative quantum field theories.

One of the key ingredients in such computations is the evaluation of multiloop scattering amplitudes, which typically involve complex algebraic calculations of multivariate rational functions. A major bottleneck in these calculations is the appearance of large expressions in the intermediate stages, which can be orders of magnitude more complicated than the final results. Thanks to the development of finite-field techniques~\cite{vonManteuffel:2014ixa, Peraro:2016wsq}, these complexities can be completely avoided. Instead, we only need to evaluate the rational functions over finite fields multiple times and reconstruct the analytic expressions at the final stage using these numerical samples. Nevertheless, it usually requires a large number of numerical samples for the reconstruction, which can be very time-consuming for cutting-edge problems.

To address this issue, people are focusing on two main approaches. The first approach aims to improve the efficiency of the generator of numerical samples. For instance, in the context of Feynman integrals reduction, which involves solving large linear systems comprised of integration-by-parts (IBP) identities~\cite{Chetyrkin:1981qh, Laporta:2001dd} to derive reduction relations for target integrals, significant progress has been made in refining the linear systems. Techniques such as syzygy equations~\cite{Gluza:2010ws, Schabinger:2011dz, Larsen:2015ped, Boehm:2018fpv, Bendle:2019csk, Wu:2023upw} and block-triangular systems among integrals~\cite{Liu:2018dmc, Guan:2019bcx} have been developed to reshape the IBP systems and improve the efficiency of numerical computation. There have also been progress in improving the solver for linear systems~\cite{Klappert:2020nbg, Magerya:2022hvj}, which also enables faster generation of numerical samples. The second approach focuses on reducing the number of samples required for reconstructing the rational functions. There have been many attempts, such as employing more effective interpolation techniques~\cite{Peraro:2016wsq, Klappert:2019emp, Peraro:2019svx, Belitsky:2023qho} and making better ansatz for the functions~\cite{Badger:2021imn, DeLaurentis:2022otd, Abreu:2023bdp}.

In this Letter, we propose a novel method that aims to further advance the second approach by finding all the independent linear relations among target rational functions. The method is inspired by a simple observation that the usual reconstruction strategy is not fully optimized when the rational functions share common structures, such as a common set of denominators, as each function is treated individually. To illustrate this point, let's consider a simple example with 100 univariate functions given by
\begin{align}\label{eq:simple}
f_i(x) = \left(\frac{1+x}{1-x}\right)^{i-1}, \quad i\in [1, 100].
\end{align}
Naively, we would need to evaluate these functions approximately 200 times in order to reconstruct them using e.g. Thiele's interpolation formula. However, by recognizing the common structures they share, we can immediately observe that they satisfy a system of 99 linear relations
\begin{align}\label{eq:simple_system}
(1-x)f_{i+1}(x) - (1+x)f_{i}(x) = 0, \quad i\in[1, 99],
\end{align}
which represents all the independent relations. This linear system is almost an equivalent form of the explicit solutions given in Eq.~\eqref{eq:simple}, because the latter can be recovered by solving the former after fixing the only degree of freedom via $f_1(x) = 1$. The surprising thing is that the polynomial degree of the coefficients in the linear system is much smaller than that of the explicit functions, resulting in much fewer required samples to construct the system. Specifically, we can make the following ansatz
\begin{align}\label{eq:simple_ansatz}
(a_i+b_ix)f_{i+1}(x)+(c_i+d_ix)f_{i}(x) = 0,
\end{align}
and fit $a_i$, $b_i$, $c_i$ and $d_i$ with at most four samples. This shows the basic philosophy of our method: by exploiting the independent linear relations among the functions, we can effectively utilize their shared structures to reduce the number of samples required.

\sect{The method}
Consider a set of $n$ unknown nonzero rational functions $\{f_1(\vec{x}), \ldots, f_n(\vec{x})\}$ that depend on $k$ variables $\vec{x} = \{x_1, \cdots, x_k\}$. Our goal is to determine all $n-1$ independent linear relations among these functions\footnote{The total number of independent relations is $n-1$. This can be proved by the following argument. First, the number should not be $n$, otherwise all the functions would vanish. Second, the number should not be less than $n-1$, otherwise more than one function can be chosen as our basis, which is impossible, since for any two functions $f_i$ and $f_j$ in the basis, there is always an identity $f_i - (f_i/f_j) \times f_j = 0$.}. In practice, it is not always possible to make a perfect ansatz like Eq.~\eqref{eq:simple_ansatz}, where the involved functions and the form of their coefficients are specified in a highly concise way, before knowing the explicit solutions of the functions. However, it is still possible to systematically propose reasonable ansatz. In general, the ansatz of a linear relation can be written as
\begin{align}\label{eq:ansatz}
Q_1(\vec{x})f_1(\vec{x})+\cdots+Q_n(\vec{x})f_n(\vec{x}) = 0,
\end{align}
where the $Q$'s are polynomials of $\vec{x}$. Since they are polynomials, it suffices to specify the monomials $x_1^{\alpha_1}\cdots x_k^{\alpha_k}$ that are permitted to appear in each $Q_i$. For convenience, we define the set of monomials up to degree $m$ as
\begin{align}
P(\{x_{i_1},...,x_{i_l}\}, m):=\{x_{i_1}^{\alpha_1}\cdots x_{i_l}^{\alpha_l}|\textstyle\sum\alpha_i\leq m\},
\end{align}
and the product between two sets of monomials as
\begin{align}
P_1\times P_2:=\{p_1 p_2\,|\,p_1\in P_1, p_2 \in P_2\}.
\end{align}
To begin, we divide the variables $\{x_1, \ldots, x_k\}$ into $r$ subsets $S_1,\ldots,S_r$. Then, given $r$ non-negative integers $\vec{z} = \{z_1, \ldots, z_r\}$, we are able to construct a set of monomials $M(\vec{z})$ for a specific $Q_i$ as
\begin{align}
M(\vec{z}):=P(S_1, z_1)\times\cdots\times P(S_r, z_r).
\end{align}
In practical applications where a priori knowledge about the functions is not available, we can construct the same set of monomials for all the $Q$'s. As a result, the ansatz~\eqref{eq:ansatz} is fully determined by the integers $\vec{z}$.

To search all the independent relations, we employ and extend the algorithm presented in Ref.~\cite{Guan:2019bcx}. We start with $z_1=\cdots=z_r=0$, which is the solution of $\sum z_i = 0$. Using this configuration, we construct the ansatz and fit the unknown coefficients to obtain a set of linear relations. This can be achieved by sampling the functions over a specific finite field and solving the resulting linear equations within that finite field. If the number of obtained relations is insufficient, we can proceed to $\sum z_i=1$, which yields $r$ solutions. For each solution, an ansatz is formulated again, and the unknown coefficients are fitted accordingly over the same finite field. If additional relations are still required, we continue to $\sum z_i=2$ and repeat the process. By iteratively increasing the value of $\sum z_i$, we can eventually obtain all the desired independent relations over a finite field.

There is a very subtle aspect to consider during this process. For two different configurations $\vec{z}$ and $\vec{z}^\prime$, it is possible that $z_i \leq z_i^\prime$ holds for all $i$. In such cases, the ansatz of $\vec{z}$ is entirely covered by the ansatz of $\vec{z}^\prime$. Therefore, the relations obtained from $\vec{z}$ would be redundantly obtained from $\vec{z}^\prime$, which is unnecessary. To address this issue, before performing the linear fit of $\vec{z}^\prime$, we manually eliminate the monomials in $Q_i$ that belong to the following set:
\begin{align}
\Lambda_i(\vec{z})\times M(\vec{z}^\prime-\vec{z}).
\end{align}
Here, $\Lambda_i(\vec{z})$ is the set of all ``solved'' monomials of $f_i$ from $\vec{z}$. A monomial is said to be ``solved'' if its coefficient is treated as a free variable in the fitting process. For example, if the following ansatz
\begin{align}
(a x_1 + b x_2+c) f_1 + (d x_1+ e x_2+g)f_2 +\cdots = 0
\end{align}
has a 2-dimensional solution space and $a$ and $e$ are set as free variables, then the monomial $x_1$ for $f_1$ and $x_2$ for $f_2$ are considered as ``solved''. This way, we effectively eliminate the redundant information from the solution space of $\vec{z}^\prime$, resulting in almost distinct relations. Furthermore, the number of unknowns of $\vec{z}^\prime$ is reduced to some extent, allowing for more efficient sampling and fitting.

Finally, to recover the explicit results using the traditional reconstruction algorithm described in Ref.~\cite{vonManteuffel:2014ixa, Peraro:2016wsq}, we typically need to extract information from additional finite fields. As we move to other finite fields, we can leverage the knowledge acquired during the aforementioned search process. Specifically, we can entirely discard the ansatz that fails to generate independent relations. For the ansatz that does yield independent relations, we can eliminate all the monomials from the original ansatz whose coefficients have been determined to be zero. This further reduces the number of samples required.

\begin{figure}[htb]
  \centering
  \includegraphics[width=0.47\textwidth]{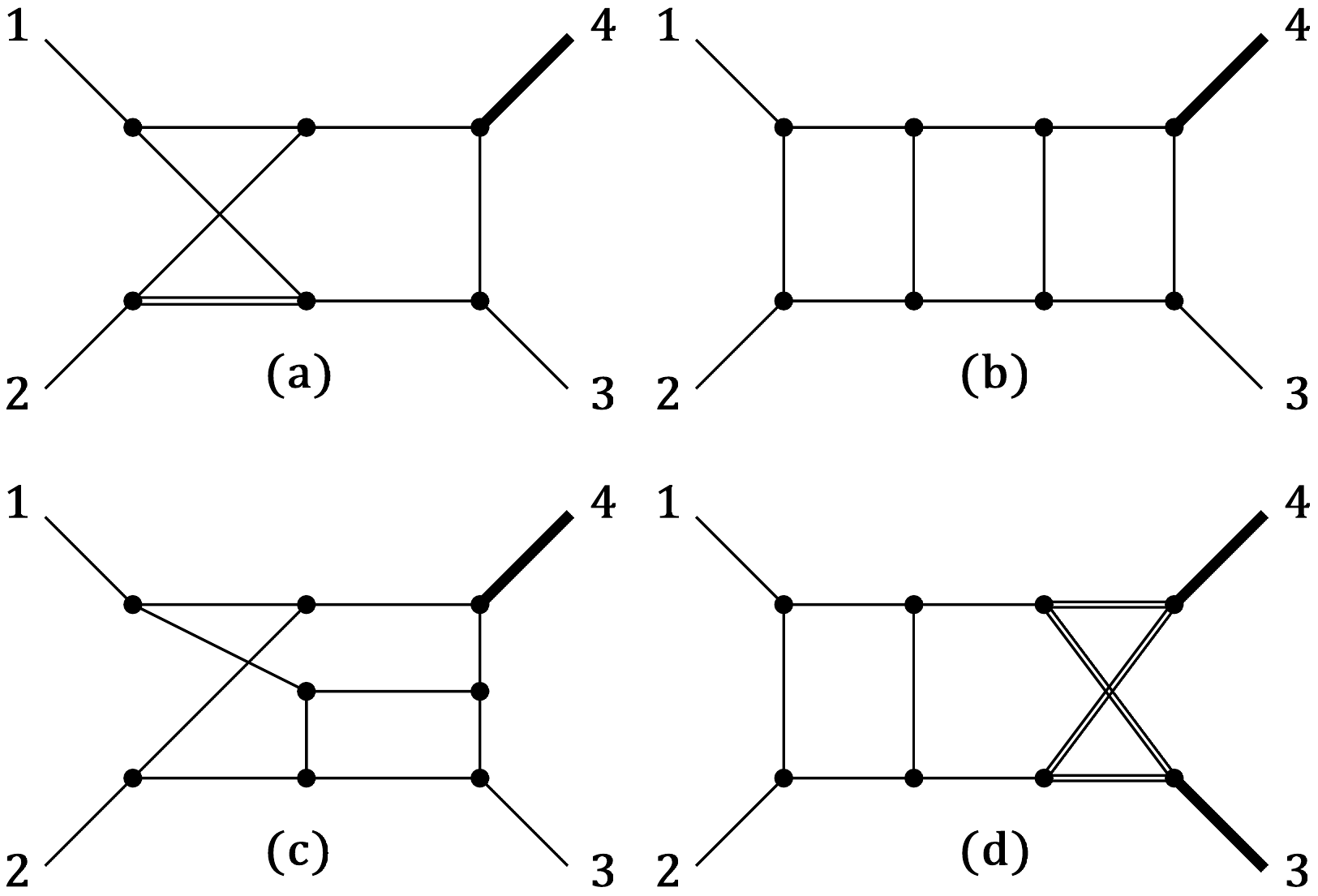}\\
  \caption{Example topologies. Bold lines and double lines represent massive particles with different masses, while the remaining solid lines represent massless particles.}\label{fig:diagram}
\end{figure}

\sect{Examples}
We demonstrate the power of our method in the context of Feynman integrals reduction, with some example topologies shown in Fig.~\ref{fig:diagram}. This involves expressing scattering amplitudes or Feynman integrals as linear combinations of a set of master integrals $\{\mathcal{M}_1,\ldots,\mathcal{M}_n\}$, such as:
\begin{align}
\mathcal{A} = f_1\mathcal{M}_1+\cdots+f_n\mathcal{M}_n,
\end{align}
where $f_1,\cdots,f_n$ are rational functions of space-time dimension $D$ and kinematic variables. If the master integrals are chosen in a standard manner, such as using the Laporta basis or canonical basis, then these functions always share a common set of denominators. These common structures can be effectively utilized to reduce the number of samples required using our method.

Throughout our examples, we consistently select the Laporta basis as master integrals and introduce an auxiliary function $f_{n+1} = 1$ for each amplitude or integral requiring reduction. This function serves as the basis for our functions, enabling us to express all the unknown functions explicitly after determining all $n$ independent relations.

The first example is taken from the two-loop amplitude of the mixed QCD-electroweak correction to $pp\to Z+j$ \cite{zjew}. We focus on a portion of the amplitude corresponding to the topology shown in Fig.~\ref{fig:diagram}~(a), which involves 56 master integrals. To be able to generate numerical samples for these 56 reduction coefficients, we employ \texttt{LiteRed}~\cite{Lee:2013mka} to construct the IBP system and \texttt{FiniteFlow}~\cite{Peraro:2019svx} to perform numerical evaluations over finite fields with 64-bit prime numbers. During the computation, the squared masses of the $Z$ boson (bold line) and the $W$ boson (double line) are set to $1$ and $7/9$, respectively. As a result, the problem depends only on three variables: dimensional regulator $\epsilon = (4-D)/2$ and two Mandelstam variables $s_{12}$ and $s_{13}$, where $s_{ij} := (p_i+p_j)^2$.

Now we can apply our method to find all 56 independent relations. We start by dividing the three variables into two subsets: $\{\epsilon\}$ and $\{s_{12}, s_{13}\}$. Next, we construct the ansatz and fit the unknowns using various configurations of $\{z_1, z_2\}$ obtained through iterative increments of $z_1+z_2$. The number of independent relations reaches 56 when $z_1+z_2 = 6$. After this, we proceed to two additional finite fields and reconstruct the explicit solutions by solving the obtained linear systems over these fields.

The details for the samples are summarized in Tab.~\ref{tab:cost}. It is evident from the table that the number of required samples over a finite field are significantly reduced, from 18326 to 2199, and further to 1561 after leveraging the knowledge gained from the first run, due to the drastic reduction in the polynomial degree. The details for the computational cost are summarized in Tab.~\ref{tab:cost2}. A remarkable observation is that our linear system can be solved with significantly higher efficiency than the IBP system over finite fields, as a result of its much smaller size. In particular, the time it takes for our system to generate a numerical sample is $7.5\times 10^{-4}$s, while the IBP system requires $0.3$s. This makes us 400 times faster. Consequently, in most cases the computational cost in our approach is dominated by the generation of the samples $T_\text{sam}$ and the improvement factor for the computational cost $R_2$ can be roughly estimated by that for the required samples $R_1$. In summary, we improve the reduction efficiency by a factor of $R_2=9.8$ in this example.

The second example comes from a three-loop four-point one-mass planar topology illustrated in Fig.~\ref{fig:diagram}~(b). We focus on a rank-6 integral in the top sector, which can be reduced to 83 master integrals using the improved IBP system provided in Ref.~\cite{Wu:2023upw}. This improved system is generated utilizing Syzygy equations~\cite{Gluza:2010ws, Schabinger:2011dz, Larsen:2015ped, Boehm:2018fpv, Bendle:2019csk} and exhibits better performance than the naive IBP system. Once again, we assume a unit mass for $p_4$ and divide the remaining variables into two subsets: $\{\epsilon\}$ and $\{s_{12}, s_{13}\}$. The system of 83 independent relations is complete when $z_1+z_2 = 8$. See Tab.~\ref{tab:cost} and \ref{tab:cost2} for more detailed information. In short, we achieve a 9.6-fold improvement this time. In this example, we also observe a significant improvement in our linear system compared to the IBP system, even though the IBP system has been refined.

Our third example depicted in Fig.~\ref{fig:diagram}~(c) is a three-loop four-point one-mass nonplanar topology. In this example, our goal is to derive the differential equations~\cite{Kotikov:1990kg, Kotikov:1991pm, Remiddi:1997ny, Gehrmann:1999as, Argeri:2007up, MullerStach:2012mp, Henn:2013pwa, Henn:2014qga} for the master integrals with respect to the Mandelstam variables $s_{12}$ and $s_{13}$. We generate the IBP system using \texttt{LiteRed}, which yields 280 master integrals and is capable of reducing their derivatives. Since there are multiple integrals that require reduction, we also have to build multiple linear systems. However, fortunately, the numerical samples can be recycled, so the total number of samples remains manageable. As usual, we set $p_4^2 = 1$ and divide the variables into $\{\epsilon\}$ and $\{s_{12}, s_{13}\}$. All the linear systems are complete before $z_1+z_2$ reaches 9. More details can be found in Tab.~\ref{tab:cost} and \ref{tab:cost2}. We note that this example is too complicated to be computed within a reasonable time using the traditional reconstruction strategy with the naive IBP system and an unrefined choice of master integrals. Therefore, the total cost $T_\text{tra}$ in Tab.~\ref{tab:cost2} is estimated by multiplying the computational cost of a single sample by the total number of required samples. In summary, the improvement in efficiency is approximately 53-fold. Notably, despite the requirement to solve multiple linear systems to obtain the final explicit solutions in this example, we still have $T_\text{sol}\ll T_\text{sam}$, thanks to the extremely high efficiency of our systems.

Our final example, as shown in Fig.~\ref{fig:diagram}~(d), is a topology relevant for NNLO correction to double Higgs production in full QCD. We aim to construct differential equations with respect to the squared top mass, while keeping other kinematic variables fixed to rational numbers. This type of problem is extensively involved in the auxiliary mass flow method~\cite{Liu:2017jxz, Liu:2020kpc, Liu:2021wks, Liu:2022chg} for the numerical computation of master integrals, where differential equations with respect to internal masses are required and can be a major bottleneck for cutting-edge problems. To begin, we utilize \texttt{LiteRed} to generate the IBP system, enabling us to derive a closed system of differential equations for 336 master integrals. We then set $s_{12}=10$, $s_{13}=-22/9$ and $m_H^2=1$, and divide the remaining two variables into $\{\epsilon\}$ and $\{m_t^2\}$. All the linear systems are successfully constructed with the condition $z_1+z_2\leq5$. The summarized details can be found in Tab.~\ref{tab:cost} and \ref{tab:cost2}. Overall, we obtain a 9.5-fold improvement. Nevertheless, it is worth noting that due to the larger number of master integrals, both the size and the number of complicated linear systems increase, making it more challenging to obtain the explicit results. Therefore, $T_\text{sol}$ and $T_\text{sam}$ have comparable values in this example.

\begin{table}[htbp]
	\centering
	\begin{tabular}{|c|c|c|c|c|c|c|c|c|}\hline
    Top. & $N_0$ & $N_1$ & $N_2$ & $N_\text{p}$ & $d_\text{num}$ & $d_\text{den}$&$d_\text{rel}$ & $R_1$\\\hline
    (a)  & 18326 & 2199 & 1561 & 1+2 & 49 & 48 &6&  10.3\\\hline
    (b)  & 48574 & 6010 & 4599  & 1+2 & 65 & 64 &8&  9.6 \\\hline
    (c)  & 391937 & 9612 & 6810 & 1+5 & 122 &123& 8  & 53.9  \\\hline
    (d)  &  14362 &  1414  &1248 & 1+32 & 145 & 144 &5&  11.5\\\hline
	\end{tabular}
	\caption{$N_0$ represents the number of samples required for reconstructing the functions over a finite field using the traditional strategy, given by \texttt{FiniteFlow}. $N_1$ corresponds to the number of samples required to find all the independent relations in the search process, while $N_2$ indicates the number of samples necessary for computations over subsequent finite fields. $N_\text{p}$ signifies the number of finite fields needed to reconstruct the rational numbers. $d_\text{num}$ and $d_\text{den}$ denote the total polynomial degree of the numerator and the denominator of the most complicated target function, respectively. $d_\text{rel}$ represents the polynomial degree of the most complicated coefficient in our linear relations. $R_1:=N_0N_\text{p}/(N_1+N_2(N_\text{p}-1))$ represents the improvement factor for the required samples.}
	\label{tab:cost}
\end{table}

\begin{table}[htbp]
	\centering
	\begin{tabular}{|c|c|c|c|c|c|c|c|c|}\hline
    Top. & $N_\text{ibp}$ & $N_\text{rel}$ & $t_\text{ibp}/\text{s}$ & $t_\text{rel}/\text{s}$ & $T_\text{tra}/\text{h}$ & $T_\text{sam}/\text{h}$ &$T_\text{sol}/\text{h}$ & $R_2$\\\hline
    (a)  & 34336 & 56 & 0.3 & 0.00075 & 4.6 & 0.44 & 0.03 & 9.8\\\hline
    (b)  & 200074 & 83 & 1.9  & 0.0024 & 78.5 & 8.03 & 0.12 & 9.6 \\\hline
    (c)  & 3461628 & 280 & 690 & 0.013 & $450728^*$ & 8369 & 180 & 53  \\\hline
    (d)  &  625070 &  336  &24.5 & 0.019 & 3230 & 281 & 59 & 9.5\\\hline
	\end{tabular}
	\caption{$N_\text{ibp}$ and $N_\text{rel}$ represent the number of linear equations in the IBP system and our linear system\footnote{For topologies (c) and (d), we choose the most complicated linear system as a representative example. Same for $t_\text{rel}$.}, respectively. $t_\text{ibp}$ and $t_\text{rel}$ correspond to the CPU time to obtain a single finite-field sample by solving the IBP system and our linear system via \texttt{FiniteFlow}, respectively. $T_\text{tra}$ denotes the total CPU time required for reconstructing the functions using the traditional strategy, while the $*$ notation means that the cost is estimated. $T_\text{sam}$ and $T_\text{sol}$ represent the total CPU time needed for generating samples and solving the systems to obtain explicit solutions in our approach, respectively. $R_2 := T_\text{tra}/(T_\text{sam}+T_\text{sol})$ denotes the improvement factor for the computational cost.}
	\label{tab:cost2}
\end{table}

All of the results have been validated by using several random numerical samples from the generators. For each example, the explicit reduction coefficients along with the linear system they satisfy, are provided in the ancillary files~\cite{examples} for interested readers to examine and test.

\sect{Summary and outlook}
In this Letter, we present a novel method for the reconstruction of rational functions, addressing one of the main bottlenecks in high-precision calculations in particle physics. By exploiting all the independent relations among functions with shared structures, our method substantially reduces the polynomial degree and thereby the number of required numerical samples. We provide cutting-edge examples in the context of Feynman integrals reduction, illustrating how our method significantly improves the computational efficiency. These advances make our approach useful for future calculations in particle physics.

In our current examples, the computational cost of determining relations among functions by solving linear equations over finite fields is not presented, as it is negligible compared to other costs. Based on our experience, this holds true when the number of unknowns (samples) is less than $\mathcal{O}(20,000)$, which is applicable to most 2-variate and 3-variate problems. However, in scenarios with more than three variables, such as in the case of two-loop five-point amplitude reduction, the use of a dense ansatz in Eq.~\eqref{eq:ansatz} may result in a substantial number of unknowns, making the determination of relations a potential major bottleneck. To address this issue, a better approach is required for generating the ansatz. This involves not only improving the ansatz for the polynomials $Q_i(\vec{x})$, but also introducing more effective auxiliary functions. For the first aspect, adopting a sparse or semi-sparse ansatz could prove beneficial. We anticipate the existence of such an ansatz, similar to the one used in the traditional strategy, where functions are initially reconstructed along one-dimensional slices, simplifying the computation process significantly. For the second aspect, we can draw inspiration from existing literature. For instance, in Ref.~\cite{Badger:2021imn}, it was observed that introducing suitable auxiliary functions facilitated the identification of linear relations at the rational number level in the context of two-loop five-point amplitude reduction. We expect that similar observations may hold for relations with higher degrees, but further investigation is required and left for future study.

It is also noteworthy that for the first three examples, the computational cost associated with solving the systems to obtain explicit solutions is inconsequential compared to the cost of generating the required samples from the IBP system. This is primarily due to the smaller size of our linear systems relative to the IBP systems. Consequently, any significant improvement in the IBP systems would yield substantial benefits for our method. In the case of the last example, these two computational costs are comparable, with the linear systems being significantly complicated. In such cases, a refined approach to grouping the target functions is necessary, and we defer this investigation to future research.

\begin{acknowledgments}
\sect{Acknowledgments}
I would like to thank Fabrizio Caola for fruitful discussions about the method and valuable suggestions on the manuscript. The work was supported by the ERC Starting Grant 804394 \textsc{HipQCD} and by the UK Science and Technology Facilities Council (STFC) under grant ST/T000864/1. {\tt JaxoDraw}~\cite{BINOSI200476} was used to generate Feynman diagrams.
\end{acknowledgments}

%


\end{document}